\documentclass[aps,prl,twocolumn,float]{revtex4}
\usepackage{amsmath,bm,epsfig}
\newcommand{\B}[1]{{\bm{#1}}}
\usepackage[dvips]{color}
\begin{document}

\title{What determines the static force chains in stressed granular media?}
\author{Oleg Gendelman$^*$, Yoav G. Pollack, Itamar Procaccia, Shiladitya Sengupta  and Jacques Zylberg}
\affiliation{Dept. of Chemical Physics, The Weizmann Institute of Science,  Rehovot 76100, Israel\\$*$
Faculty of Mechanical Engineering, Technion, Haifa 32000, Israel}
\begin{abstract}
The determination of the normal and transverse (frictional) inter-particle forces within a granular medium is a long standing, daunting, and yet unresolved problem. We present a new formalism which employs the knowledge of the external forces and
the orientations of contacts between particles (of any given sizes), to compute all the inter-particle forces.
Having solved this problem we exemplify the efficacy of the formalism showing that the force chains in such systems are determined by an
expansion in the eigenfunctions of a newly defined operator.
\end{abstract}
\maketitle
\begin{figure}
\includegraphics[scale = 0.15]{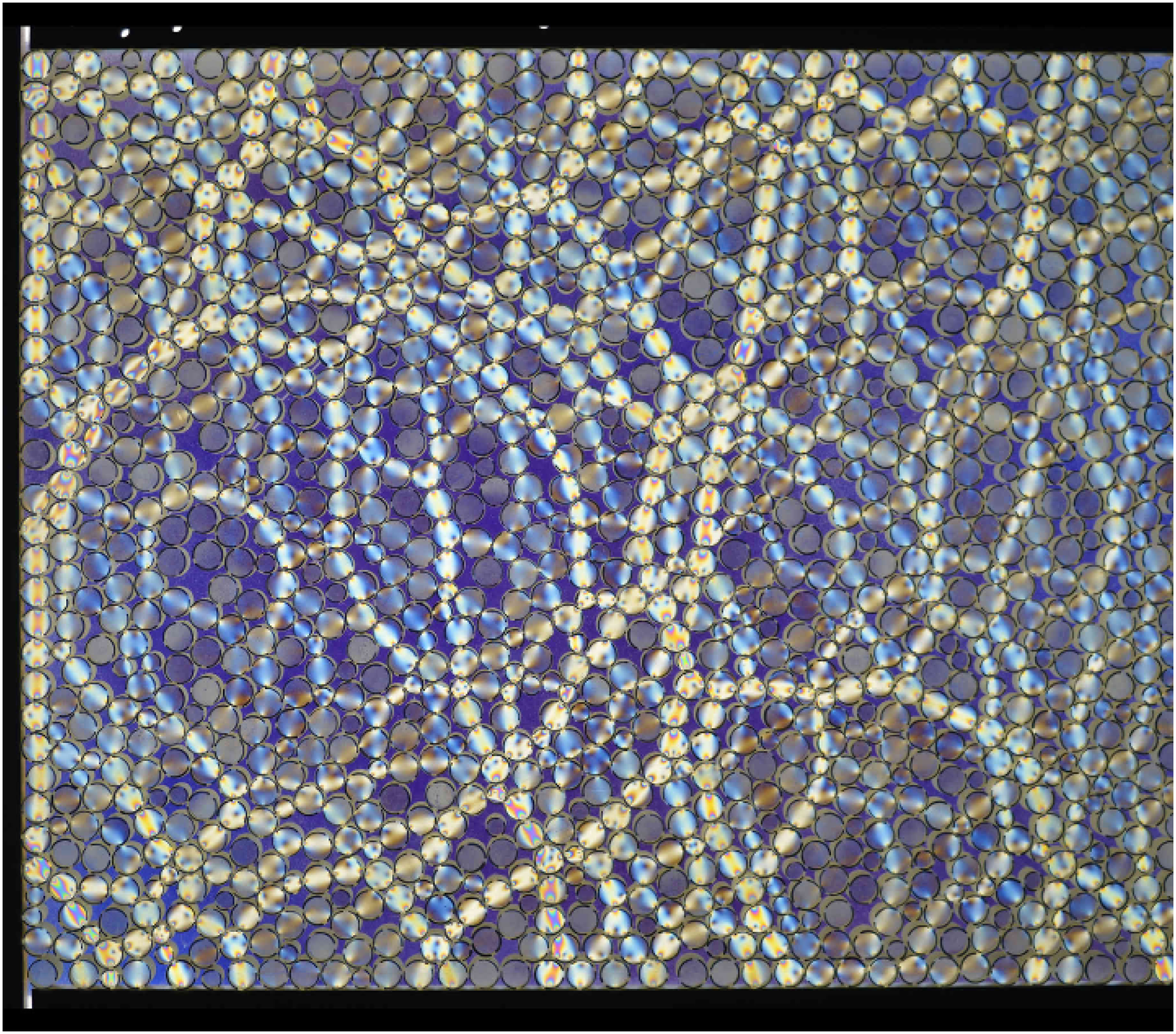}
\caption{Force chains in a binary system of plastic disks of two diameters, stressed uniaxially at the boundaries. The force chains are
made visual by the optical birefringence of the stressed disks. The image
is courtesy of V. Sathish Akella and Mahesh Bandi, Okinawa Institute of Science and Technology.}
\label{mahesh}
\end{figure}

In a highly influential paper from 2005 Majmudar and Behringer \cite{05MB} wrote: ``Inter-particle forces in granular media form an inhomogeneous distribution of filamentary force chains. Understanding such forces and their spatial correlations, specifically in response to forces at the system boundaries, represents a fundamental goal of granular mechanics. The problem is of relevance to civil engineering, geophysics and physics, being important for the understanding of jamming, shear-induced yielding and mechanical response." A visual example
of such force chains in a system of plastic disks is provided in Fig.~\ref{mahesh}. In this Letter
we present a solution of this goal.

To be precise, the problem that we solve is the following: consider a granular medium with known sizes of the granules,
for example the 2-dimensional systems analyzed in Ref.~\cite{05MB} and shown in Fig.~\ref{mahesh}, of $N$ disks of known diameters $\{\sigma_i\}_{i=1}^N$. Given the external forces, denoted below as $\B F^{\rm ext}_i$ and the external torques $\Gamma^{\rm ext}_i$ exerted on the granules, and given
the angular orientations of the vectors connecting the center of masses of contacting granules (but not the distance between them!), determine all the inter-particle normal and tangential forces $\B f^n_{ij}$ and  $\B f^t_{ij}$. The method presented below applies to granular systems in mechanical equilibrium; the issue of instabilities and abrupt changes in the force chains will be discussed elsewhere. For the sake of clarity and
simplicity we will present here the two-dimensional case; the savvy reader will recognize that the formalism
and the solution presented will go smoothly also for the three-dimensional case (as long as the system is in mechanical
equilibrium). The full formalism will be presented in a longer publication in due course.

The obvious conditions for mechanical equilibrium are that the forces and the torques on each particle have to sum up to zero \cite{98Ale}. The condition of force balance is usefully presented in matrix form using the following notation. Denote the (signed) amplitudes of the inter-particle forces $f_{ij}$ as a vector $|f\rangle$, where the amplitudes  $f^n_{ij}$ appear first and then the amplitudes $f^t_{ij}$. The vector of $x$ and $y$ components $F^{\rm ext}_{i,x}$ and $F^{\rm ext}_{i,y}$ is denoted as $|F^{\rm ext}\rangle$ where all the $x$ components
are presented in $|F^{\rm ext}\rangle$ first and then all the $y$ components. The vector $|f\rangle$ has $2c$ entries where $c$ is the number of contacts between particles. The vector $|F^{\rm ext}\rangle$ has $2N$ entries where $N$ is the number of particles, with zero entries for all the particles on which there is no external force.  We can then write the force balance condition as
\begin{equation}
M|f\rangle = - |F^{\rm ext}\rangle \ ,
\label{M}
\end{equation}
where $M$ is a $2N\times 2c$ matrix. The entries in the matrix $M$ contain the directional information, see Supplemental Material at [URL will be inserted by publisher] for an example of an $M$ matrix. Denote the unit vector
 in the direction of the vector distance between the centers of mass of particles $i$ and $j$ by $\B {\hat n}_{ij}$, and the tangential
 vector by $\B {\hat t}_{ij}$ orthogonal to $\B {\hat n}_{ij}$. Then the entries of $M$
 display the projections $\hat n_{ij,x}$ and $\hat n_{ij,y}$ or ${\hat t}_{ij,x}$ and ${\hat t}_{ij,y}$ as appropriate. We thus guarantee that Eq.(\ref{M}) is equivalent to the mechanical equilibrium condition
\begin{equation}
\sum_j \B f_{ij}+\B F^{\rm ext}_i =0\ .
\label{sum}
\end{equation}
As is well known, the friction-less granular system in the thermodynamic limit is jammed exactly at the isostatic condition $2N=c$ \cite{05WNW}. In the friction-less case $M$ is a $2N\times c$ matrix and as long as $c=2N$ one can solve the problem by multiplying Eq.~(\ref{M}) by the transpose $M^T$, getting
\begin{equation}
M^T M|f\rangle = - M^T|F^{\rm ext}\rangle \ .
\label{MMT}
\end{equation}
In this case the matrix $M^T M$ has generically exactly three Goldstone modes (two for translation and one for rotation) \cite{foot}, and since the external
force vector is orthogonal to the Goldstone modes (otherwise the external forces will translate or rotate the system), Eq.~(\ref{MMT}) can
be inverted with impunity by multiplying by $[M^T M]^{-1}$.  In fact even when $c<2N$ but the system is small enough to be jammed, this method can be used since there are enough constraints to solve for the forces. This last comment is important for our developments below.

The problem becomes under-determined above isostaticity in the frictionless case, when force chains begin to build up that span from one boundary to the other. With friction we anyway have twice as many unknowns and we need to add the constrains of torque balance.
The condition of torque balance is $\sum_j \B r_i\times \B f^t_{ij}+\B \Gamma_i^{\rm ext}=0$ on every particle, where $\B \Gamma_i^{\rm ext}$ is the external torque exerted on the $i$th disk \cite{foot}. For disks, $\B r_i$ is in the normal direction, and therefore the torque balance
becomes a condition that the sum of tangential forces has to balance the external tangential force.  This condition can be added to Eq.~(\ref{M}) using a new matrix $\B B$ in the form
\begin{equation}
B|f\rangle \equiv
\begin{pmatrix}
  &&M&& \\
  0 && ~ && T
\end{pmatrix}\left|
\begin{pmatrix}
  f^n \\
  f^t\\
\end{pmatrix}\right\rangle =
-\left|
\begin{pmatrix}
 F^{\rm ext}_{x} \\
  F^{\rm ext}_{y} \\
  \Gamma^{\rm ext}
\end{pmatrix}\right\rangle \ .
\label{MT}
\end{equation}
The order of the extended matrix $\B B$ is $3N\times 2c$, see Supplemental Material at [URL will be inserted by publisher] for an example of $\B T$. Above jamming when the number of
contacts increases $2c\gg 3N$. The matrix $B$ is not square, and the matrix $B^T B$ which is
of size $2c\times 2c$, has at least $2c-3N$ zero modes \cite{footnote}. Accordingly it cannot be inverted and one can conclude that {\bf the conditions of
mechanical equilibrium are not sufficient to determine all the forces.}

\begin{figure}[h!]
\includegraphics[scale = 0.55]{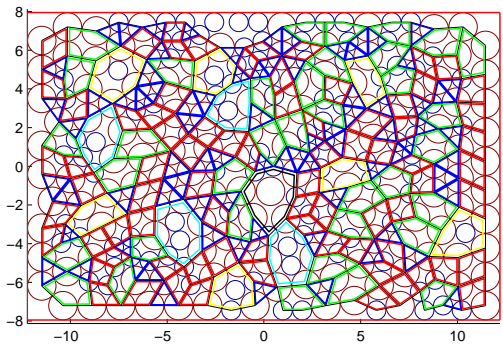}
\caption{A generic situation in a stressed arrangement of $N=242$ binary disks of diameters $\sigma_1=1$, $\sigma_2=1.5$. Here $c=432$. Shown are the $P=205$ polygons: 80 triangles in blue, 77 squares in red, 36 pentagons in green, 7 hexagons in yellow, 4 heptagons in cyan and 1 octagon in black. The Euler characteristic of the system is as expected $N - c + (P+1) = 2+R$
 since we have 14 rattlers.}
\label{geometry}
\end{figure}

Obviously what is missing are additional constraints to remove the host of zero modes. These additional  constraints are {\em geometrical} constraints \cite{02BB,04Blu} which can be read from those disks which describe connected polygons. In other words, since we know the orientation $\B {\hat n}_{ij}$ of each contact in our system, we can determine which granules are stressed in a triangular arrangement, and which in a square or pentagonal etc., see Fig.~\ref{geometry}.
Each such arrangement is a constraint on the radius vectors adjoining the centers of mass. For example if particles $i, j$ and $k$ are in
a triangular arrangement then $\B r_{ij}+\B r_{jk}+\B r_{ki}=0$, with the analogous constraint on squares, pentagons etc. These constraints can be written in a matrix form by denoting the {\em amplitudes} of inter-particle vector distances as $|r\rangle$ where we again arrange the $x$ components first and the $y$ components second:
\begin{equation}
Q|r\rangle =0 \ ,
\label{Q}
\end{equation}
where the matrix $Q$ again has entries $\hat n_{ij,x}$ or $\hat n_{ij,y}$ as appropriate to represent the vectorial geometric constraints, see Supplemental Material at [URL will be inserted by publisher] for an example of $\B Q$. Denoting the total number of polygons by $P$ the dimension of the matrix $Q$ is $2P\times c$. Of course $|r\rangle$ has $c$ entries while $|f\rangle$ had $2c$ entries. Note that in generic situations there can be also disks which are not stressed at all.
These are referred to as ``rattlers". For example in the configuration shown in Fig.~\ref{geometry} there exist 14 rattlers.

At this point we specialize the treatment to Hookean normal forces with a given force constant $\kappa$ \cite{07MSLB}. Non Hookean forces result in a nonlinear theory that can still be solved but much less elegantly. For the present case
\begin{equation}
\B f^n_{ij} = \kappa [(\sigma_i+\sigma_j)\B{\hat n}_{ij}/2-\B r_{ij}] \ .
\label{sig}
\end{equation}
Denoting the amplitudes of the vectors $(\sigma_i+\sigma_j)\B {\hat n}_{ij}/2$ as the vector $|\sigma\rangle$ (again with first the $x$ and then
the $y$ components), we can rewrite Eq.~(\ref{Q}) in the form
\begin{eqnarray}
Q|r\rangle &=&Q|\sigma-f^n/\kappa\rangle=0\ , \\
Q|f^n\rangle &=& Q|\kappa \sigma\rangle \ .
\label{Q2}
\end{eqnarray}
Having this result at hand we can formulate the final problem to be solved. Arrange now a new matrix, say G, operating on a vector $|f\rangle$, with a RHS being a vector, say $|t\rangle$, made of a stacking of $|F^{\rm ext}\rangle$, $|\Gamma^{\rm ext}\rangle$ and
$Q|\kappa \sigma\rangle$, as before with $x$ and then $y$ components:
\begin{equation}
G|f\rangle \equiv
\begin{pmatrix}
  &&M&& \\
  0 && ~ && T \\
  Q && ~ && 0\\
\end{pmatrix}\left|
\begin{pmatrix}
  f^n \\
  f^t\\
\end{pmatrix}\right\rangle =
\left|
\begin{pmatrix}
  f_x^{\rm ext} \\
  f_y^{\rm ext} \\
  \Gamma^{\rm ext} \\
  Q|\kappa\sigma\rangle\\
\end{pmatrix}\right\rangle
\label{final}
\end{equation}
Using these objects our problem is now
\begin{equation}
G|f\rangle = |t\rangle \ .
\end{equation}
The dimension of the matrix $G$ is $(3N+2P)\times 2c$ and the matrix $G^T G$ has the dimension $2c\times 2c$. We can use now the Euler characteristic \cite{01Mat} to show that the situation has been returned here to the analog of the invertible matrix $M^TM$ when $c\le 2N$: the Euler characteristic in two dimensions requires that
\begin{equation}
N-c+(P+1) =2+R \ , 
\label{euler}
\end{equation}
where $R$ is the number of ``rattlers" i.e. disks on which there is no force. 
Accordingly we find that
\begin{equation}
2c=2N+2P-2-2R\ll 3N+2P \ .
\end{equation}
Consequently, the matrix $G^TG$ has no zero eigenmodes. Thus the final solution for the forces can be obtained as
\begin{equation}
|f\rangle = (G^T G)^{-1} G^T|t\rangle =\sum_i \frac{\langle \Psi_i|G^T|t\rangle}{\lambda_i }|\Psi_i\rangle \ .
\label{final}
\end{equation}
where $\Psi_i$ is the set of eigenfunctions of $G^TG$ associated with eigenvalues $\lambda_i$.
We compared the inter-particle forces obtained from direct numerical simulations (see below for details) to those computed
from Eq.~(\ref{final}). Both normal and tangential forces are of course identical to machine accuracy.
We reiterate that we did not need to know the distances between particles. This is important in applying the formalism to experiments since it is very difficult to measure with precision the degree of compression of hard particles like, say, metal balls or sand particles. Note also the remarkable
fact that we never had to provide the frictional (tangential) force law in the formalism to obtain the correct forces!

\begin{figure}[h!]
\includegraphics[scale = 0.50]{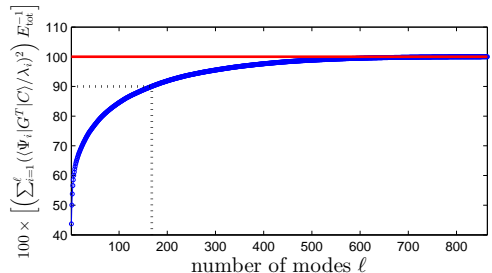}
\caption{The cumulative percentage contribution to the energy of the eigenfunctions $\Psi_i$ of $G^TG$, ordered according to the magnitude of  $\langle \Psi_i|G^T|t\rangle/\lambda_i$. The convergence is relatively fast with the first 168 leading eigenfunction (out of 864 modes) contributing 90\% of the total energy.}
\label{order}
\end{figure}

\begin{figure}[h!]
\includegraphics[scale = 0.65]{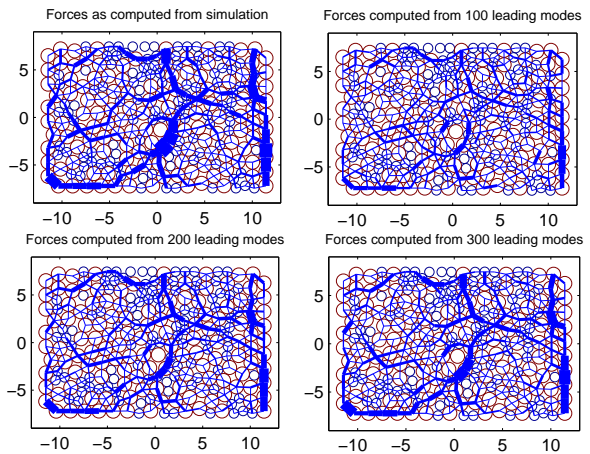}
\caption{Upper panel: the force chains in the configuration shown in Fig.~\ref{geometry}. Lower panel: the force chains as predicted from 100, 200 and 300 (energy) leading modes. While the main contributions to the force chains are visible already with 100 out of the 864 modes, the full structure is apparent only with 200. Using 300 modes we see a reasonably faithful reproduction of the force chains.}
\label{chains}
\end{figure}

At this point we can discuss the force chains. By definition these are the large forces in the system that provide the tenuous
network that keeps the system rigid. Observing Eq.~(\ref{final}) we should focus on the eigenfunction $\Psi_i$ of $G^TG$ that have
the smallest eigenvalues and the largest overlaps with $G^T|t\rangle$. These can be found and arranged in order of the magnitude of $\langle \Psi_i|G^T|t\rangle/\lambda_i$ independently of the
calculation of $|f\rangle$. In Fig.~\ref{order} we show the contribution to the total energy $\langle f|f\rangle/\kappa$, learning that about 20\% of the leading eigenfunction are responsible for 90\% of the energy.
We can therefore hope that the force chains will be determined by the same relatively small number of eigenfunctions. This is not guaranteed; due to contributions to the forces that oscillate in sign the convergence can be much slower than in the case of the energy
where the sum is of positive contributions. In Fig.~\ref{chains} we show in upper left panel the force chains in the configuration of Fig.~\ref{geometry}. In the other panels we show the prediction of the force chains using 100, 200 and 300 of the (energy) leading modes. We learn that with 100 out of the 864 modes the main force chains begin to be visible. With 200 out of the 864 modes the full structure of the force chains is already apparent, although with 300 it is represented even better.

Since the number of geometric constraints is very large,  one can ask whether all the geometric constraints are necessary, as
Eq. (\ref{euler}) shows that $2c\ll 3N+2P$. The answer is no, we could leave out constraints as long as we have enough conditions
to determine the solution. There is the obvious question then why do we have a unique solution when the number of equations is larger
than the number of unknowns. The answer to this question lies in the properties of the vector $|t\rangle$ and the matrix $G G^T$ which
does have many zero modes. A condition for the existence of a solution is that $|t\rangle$ is orthogonal to all the zero modes
of $G G^T$, as can be easily checked. We have ascertained in our simulations that this condition is always met.

In the near future we will present an extension of this formalism
to three dimensions and the use of the formalism to study the instabilities of the force networks to changes in the external
forces. As a final comment we should note that in fact only {\em one} external force is necessary to determine {\em all}
the inter-disk forces. This single external force is necessary to remove the re-scaling freedom that this problem has
by definition. 

{\bf Simulations}: For the numerical experiments in 2-dimensions we use disks of two diameters, a `small' one with diameter $\sigma_s=1.0$ and a `large' one with diameter $\sigma_\ell=1.5$. Such $N$ disks are put between
virtual walls at $x=\pm a$ and $y=\pm b$. These walls exert external forces on the disks. The external forces are taken as Hookean for simplicity. For disks near the wall at $x=\pm a$ we write
\begin{eqnarray}
F^{ext}(r_{i,x}) &=& -(r_{i,x} - a) \quad \mbox{if} \, r_{i,x} > a \nonumber\\
&=& -(r_{i,x} + a) \quad \mbox{if} \, r_{i,x} < - a \nonumber\\
&=& 0 \quad \mbox{otherwise} \ .
\label{external}
\end{eqnarray}
Here $r_{i,x}$ denotes the $x$ component of the position vector $\B r_i$ of the center of mass of the $i$th disk, and we have a similar equation for the $y$ components with $a$ replaced by $b$.
When two disks, say disk $i$ and disk $j$ are pressed against each other we define their amount of compression as $\delta r_{ij}$:
\begin{equation}
\delta r_{ij}=\sigma_{ij}-r_{ij} \ , \quad \sigma_{ij}\equiv (\sigma_i+\sigma_j)/2 \ ,
\end{equation}
where $r_{ij}$ is the actual distance between the centers of mass of the disks $i$ and $j$.

In our simulations the normal force between the disks acts along the radius vector connecting the centers of mass. We employ a Hookean force $\B f_{ij}^n=\kappa \delta r_{ij}\B {\hat n}_{ij}$.

To define the tangential force between the disks we consider (an imaginary) tangential spring at every contact which is put at rest whenever a contact between the two disks is formed. During the simulation we implement memory such that for each contact we store the signed distance $\delta t_{ij}$ to the initial rest state. For small deviations we require a linear relationship between the displacement and the acting tangential force. This relationship breaks when the magnitude of the tangential force reaches $\mu f_{ij}^n$ where due to Coulomb's law the tangential loading can no longer be stored and is thus dissipated.  When this limit is reached the bond breaks and after a slipping event the bond is restored with a the tangential spring being loaded to its full capacity (equal to the Coulomb limit).

The numerical results reported above were obtained by starting with $N=242$ particles on a rectangular grid (ratio 1:2) with small random deviations in space and no contacts. We implement a large box that contains all the particles. The box acts on the system by exerting a restoring harmonic normal force as described in Eq.~(\ref{external}). The experiment is an iterative process in which we first shrink the containing box infinitesimally (conserving the ratio). The second step is to annul all the forces and torques, to bring the system back to a state of mechanical equilibrium. We therefore annul the forces using a conjugate gradient minimizer acting to minimize the resulting forces and torque on all particles. We iterate these two steps until the system is compressed to the desired state.

\acknowledgments
This work had been supported in part by an ``ideas" grant STANPAS of the ERC. We thank Deepak Dhar for some
very useful discussions. We are grateful to Edan Lerner for reading an early version of the manuscript with very useful
remarks.



\appendix

\section*{Supplementary Material: Examples of the $M$,$T$ and $Q$ matrices}
We consider a two-dimensional configuration of $N$ particles with $C$ contacts
and $P$ polygons. For convenience of notation, only single digit particle indices are used
in this example, so that the notation $\hat{n}_{13}{}^x$ means the Cartesian $x$
component of the unit vector from the center of particle $1$ to that of particle
$3$.\\  
\begin{figure}[h!]
	\includegraphics[scale=0.25]{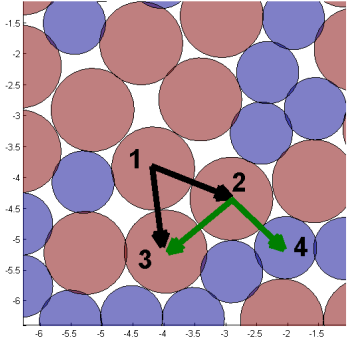}
	\caption{A subset of the particle configuration of the system. The particles' diameters in the figure are scaled-up to visually stress their finite overlap.
		Some particle indices used in the $M$ and $T$ matrices are shown. Arrows represent the normal vectors used to construct the $M$ and $T$ matrices (before normalization). Different arrow colors are for visualization purposes only.}
	\label{fig:M_configuration}
\end{figure}
The convention for ordering of the contacts is demonstrated in Eq. \ref{eq:C} (and see also Fig. \ref{fig:M_configuration}):
\begin{equation}
\left(
\begin{array}{ccccccc}
c= & 1 & 2 & \text{...} & 7 & 8 & \text{...} \\
& 1\text{with} 2 & 1 \text{with} 3 & \text{...} & 2\text{with} 3 & 2\text{with} 4 & \text{...} \\
\end{array}
\right)
\label{eq:C}
\end{equation}
The $M$ matrix is used to describe the force balance condition (Eq. 1 in the main text) and has dimension
$2N \times 2C$ in the most general case when contact forces have both normal and tangential components.
Each row is associated with a given particle $i$ and each column describes one contact and has non-zero entries
corresponding only to the pair of particles $i$ and $j$ forming that contact.
Its first $N$ rows store the $x$ components and the next
$N$ rows store the $y$ components of unit normal vectors $\hat{n}_{ij}$ and unit
tangential vectors $\hat{t}_{ij}$ (counter-clockwise orthogonal to
$\hat{n}_{ij}$). The first $C$ columns of $M$ correspond to the normal
directions and the next $C$ columns correspond to the tangential directions (which can also of course be expressed using the normal directions via a simple rotation transformation).  An
example of some of the terms of the $M$ matrix for the configuration of Fig. \ref{fig:M_configuration} is
given in Eq. \ref{eq:M}:
\begin{widetext}

	\begin{equation}
	M=\left(
	\begin{array}{cccccccccccc}
	-\hat{n}_{12}{}^x & -\hat{n}_{13}{}^x & \text{...} & 0 & 0 & \text{...}\text{...} & \hat{t}_{12}{}^x & \hat{t}_{13}{}^x & \text{...} & 0 & 0 & \text{...} \\
	\hat{n}_{12}{}^x & 0 & \text{...} & -\hat{n}_{23}{}^x & -\hat{n}_{24}{}^x & \text{...}\text{...} & -\hat{t}_{12}{}^x & 0 & \text{...} & \hat{t}_{23}{}^x & \hat{t}_{24}{}^x & \text{...} \\
	0 & \hat{n}_{13}{}^x & \text{...} & \hat{n}_{23}{}^x & 0 & \text{...}\text{...} & 0 & -\hat{t}_{13}{}^x & \text{...} & -\hat{t}_{23}{}^x & 0 & \text{...} \\
	0 & 0 & \text{...} & 0 & \hat{n}_{24}{}^x & \text{...}\text{...} & 0 & 0 & \text{...} & 0 & -\hat{t}_{24}{}^x & \text{...} \\
	: & : &   & : & : &   & : & : &   & : & : &   \\
	-\hat{n}_{12}{}^y & -\hat{n}_{13}{}^y & \text{...} & 0 & 0 & \text{...}\text{...} & \hat{t}_{12}{}^y & \hat{t}_{13}{}^y & \text{...} & 0 & 0 & \text{...} \\
	\hat{n}_{12}{}^y & 0 & \text{...} & -\hat{n}_{23}{}^y & -\hat{n}_{24}{}^y & \text{...}\text{...} & -\hat{t}_{12}{}^y & 0 & \text{...} & \hat{t}_{23}{}^y & \hat{t}_{24}{}^y & \text{...} \\
	0 & \hat{n}_{13}{}^y & \text{...} & \hat{n}_{23}{}^y & 0 & \text{...}\text{...} & 0 & -\hat{t}_{13}{}^y & \text{...} & -\hat{t}_{23}{}^y & 0 & \text{...} \\
	0 & 0 & \text{...} & 0 & \hat{n}_{24}{}^y & \text{...}\text{...} & 0 & 0 & \text{...} & 0 & -\hat{t}_{24}{}^y & \text{...} \\
	: & : &   & : & : &   & : & : &   & : & : &   \\
	\end{array}
	\right)
	\label{eq:M}
	\end{equation}
	
\end{widetext}
The $T$ matrix is used to describe the torque balance condition (see Eq. 9 in the main text) and is of
dimensions $N \times C$. Again, the row indices correspond to particles and the column
indices refer to contacts. The non-zero entries in each column correspond to the
radii of particles $i$ and $j$ forming that contact. An example of some of the
terms of the $T$ matrix for the configuration of Fig. \ref{fig:M_configuration} is given in Eq. 
\ref{eq:T}:  
\begin{equation}
T=\left(
\begin{array}{cccccc}
R_1 & R_1 & \text{...} & 0 & 0 & \text{...} \\
R_2 & 0 & \text{...} & R_2 & R_2 & \text{...} \\
0 & R_3 & \text{...} & R_3 & 0 & \text{...} \\
0 & 0 & \text{...} & 0 & R_4 & \text{...} \\
: & : &   & : & : &   \\
\end{array}
\right)
\label{eq:T}
\end{equation}
When the external torque is zero, as in our loading protocol using compression, the radii are eliminated from the torque balance equation and the  $T$ matrix can be further simplified to the form of Eq. \ref{eq:T_alt} : 
\begin{equation}
T=\left(
\begin{array}{cccccc}
1 & 1 & \text{...} & 0 & 0 & \text{...} \\
1 & 0 & \text{...} & 1 & 1 & \text{...} \\
0 & 1 & \text{...} & 1 & 0 & \text{...} \\
0 & 0 & \text{...} & 0 & 1 & \text{...} \\
: & : &   & : & : &   \\
\end{array}
\right)
\label{eq:T_alt}
\end{equation}
The $Q$ matrix (cf. Eq. 7 in the main text) is used to describe the presence of closed polygons formed by
particles in contact and and is of
dimensions $2P \times C$. Here row indices correspond to
polygons and column indices refer to the contacts. Non-zero entries in each row
describe the unit normal directions joining two particles in contact which are members
of a given polygon. The first $P$ rows store the $x$ components and the next $P$ rows
store the $y$ components of unit vectors $\hat{n}_{ij}$. An example for some of
the terms of the $Q$ matrix is given in Eq. \ref{eq:Q} (and see Fig. \ref{fig:Q_configuration}): 
\begin{equation}
Q=\left(
\begin{array}{cccccccc}
\hat{n}_{12}{}^x & -\hat{n}_{13}{}^x & 0 & \text{...} & \hat{n}_{23}{}^x & \text{...} & 0 & \text{...} \\
0 & \hat{n}_{13}{}^x & -\hat{n}_{15}{}^x & \text{...} & 0 & \text{...} & \hat{n}_{35}{}^x & \text{...} \\
: & : & : &   & : &   & : &   \\
\hat{n}_{12}{}^y & -\hat{n}_{13}{}^y & 0 & \text{...} & \hat{n}_{23}{}^y & \text{...} & 0 & \text{...} \\
0 & \hat{n}_{13}{}^y & -\hat{n}_{15}{}^y & \text{...} & 0 & \text{...} & \hat{n}_{35}{}^y & \text{...} \\
: & : & : &   & : &   & : &   \\
\end{array}
\right)
\label{eq:Q}
\end{equation}

\begin{figure}[h!]
	\includegraphics[scale=0.25]{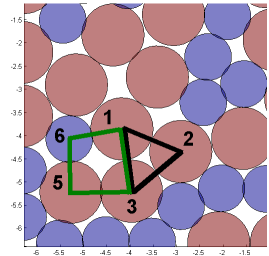}
	\caption{Same configuration as in Fig. \ref{fig:M_configuration}, only now the polygons used in the $Q$ matrix are demonstrated. }
	\label{fig:Q_configuration}
\end{figure}


\begin{thebibliography}{99}
\bibitem{foot}
The existence of ``rattlers" that are not in close contact with other particles may increase the number of zero modes.
\bibitem{footnote}
To see this note that the matrices $\B B^T\B B$ and $\B B\B B^T$ have the same rank, but $\B B\B B^T$
cannot have more than $3N$ non-zero eigenmodes. Therefore $\B B^T\B B$ must have at least $2c-3N$ zero modes.
\end{thebibliography}
\end{document}